\documentclass[12pt]{article}
\usepackage{latexsym, amsmath, epsfig}
\DeclareFontFamily{OT1}{rsfs10}{}
\DeclareFontShape{OT1}{rsfs10}{m}{n}{ <-> rsfs10 }{}
\DeclareMathAlphabet{\mathscript}{OT1}{rsfs10}{m}{n}

\if@twoside\oddsidemargin25mm
\else\oddsidemargin25mm\evensidemargin25mm\marginparwidth25mm\fi
\def\bpl{\Big(}
\def\bpr{\Big)}

\def\brr{\begin{eqnarray}}
\def\err{\end{eqnarray}}
\def\ba{\left(\begin{array}}
\def\ea{\end{array}\right)}

\newcommand{\dr}{\raise.3ex\hbox{$\stackrel{\leftarrow}{\partial }$}{}}
\newcommand{\dl}{\raise.3ex\hbox{$\stackrel{\rightarrow}{\partial}$}{}}

\newcommand{\ft}[2]{{\textstyle\frac{#1}{#2}}}
\newcommand{\ns}{\normalsize}

\renewcommand{\a}{\alpha}
\renewcommand{\b}{\beta}

\begin{document}


\begin{titlepage}

\vspace{-4cm}

\title{
   \hfill{\ns CU-TP-988\,,\,HU-EP 00/47\,,\,UPR-909T\\}
   \hfill{\ns hep-th/0011031\\[1cm]}
   {\LARGE Twisted Sectors and Chern-Simons Terms \\[.1in]
    in $M$-Theory Orbifolds\\[.5cm]  }}

\author{{\bf
   Michael Faux$^{1}$ \,,\,
   Dieter L{\"u}st$^{2}$ \,
   and Burt A.~Ovrut$^{3}$}\\[5mm]
   {\it $^1$Departments of Mathematics and Physics} \\
   {\it Columbia University} \\
   {\it 2990 Broadway, New York, NY 10027} \\[3mm]
   {\it $^2$Institut f\"ur Physik, Humboldt Universit\"at} \\
   {\it Invalidenstra\ss{}e 110, 10115 Berlin, Germany} \\[3mm]
   {\it $^3$Department of Physics, University of Pennsylvania} \\
   {\it Philadelphia, PA 19104--6396, USA}}
\date{}

\maketitle

\begin{abstract}
\noindent
It is shown that the twisted sector spectrum, as well as the associated Chern-Simons
interactions, can be determined on $M$-theory orbifold fixed planes that do
not admit gravitational anomalies. This is demonstrated for the seven-planes
arising within the context of an explicit ${\bf R \rm}^{6} \times S^{1}/
{\bf Z \rm}_{2} \times T^{4}/{\bf Z \rm}_{2}$ orbifold, although the results
are completely general. Local anomaly cancellation in this context
is shown to require fractional anomaly data that can only arise from
a twisted sector on the seven-planes, thus determining the twisted spectrum up
to a small ambiguity. These results open the door to the
construction of arbitrary $M$-theory orbifolds, including those containing
fixed four-planes which are of phenomenological interest.

\vspace{.3in}
\noindent
\end{abstract}

\thispagestyle{empty}

\end{titlepage}


\section{Introduction}

In their fundamental work \cite{hw1,hw2}, Ho\v rava and Witten discussed the
eleven-dimensional realization of $M$-theory on the background spacetime
${\bf R \rm}^{10} \times S^{1}/{\bf Z \rm}_{2}$. A main thesis in that
work is that local anomaly freedom fixes the ``twisted sector''
gauge field and matter spectrum
on each of the two ten-dimensional orbifold fixed planes almost uniquely. This
result is important in that, since the fundamental theory underlying
$M$-theory remains unknown, there is no other way to determine this
spectrum. It was subsequently demonstrated in a series of papers \cite{losw1} that if
the Ho\v rava-Witten theory is compactified on elliptically fibered Calabi-Yau
threefolds with holomorphic instantons, a quasi-realistic three family theory
of particle physics with the standard gauge group emerges. Hence, $M$-theory
orbifolds might provide a first principles foundation for low energy particle
physics.

With this in mind, it is of interest to ask whether one can construct other
orbifolds of $M$-theory beyond the $S^{1}/{\bf Z \rm}_{2}$ example of
\cite{hw1,hw2}. A
first step in this direction was taken by Dasgupta and Muhki \cite{dasmuk}
and Witten \cite{wittens5}
who discussed both local and global anomaly cancellation within the context of
$T^{5}/{\bf Z \rm}_{2}$ orbifolds. A major generalization of these results was
presented in \cite{mlo,phase,new} and \cite{bds,ksty} where all the
$M$-theory orbifolds associated with the
spacetime ${\bf R \rm}^{6} \times K3$ were constructed. In these papers, the
complete anomaly polynomials for arbitrary gravity, gauge and matter
supermultiplets were given for both ten and six-dimensional fixed planes.
Local anomaly free solutions were specified on both ten-planes and six-planes
and these were ``woven'' together to form orbifolds that were, in addition,
free of global anomalies. Furthermore, it was shown in \cite{phase,new}
that many of these
orbifold solutions were simply related to each other through the emission and
absorption of $M$-five-branes at the six-planes, through the process of
``small instanton'' phase transitions \cite{opp}.

In this paper we would like to address an essential physical
feature of 
$M$-theory orbifolds, namely:
{\it how is it possible on orbifold
fixed planes that do not admit a gravitational anomaly to determine the
twisted sector spectrum}? Recall that the main feature of the original
Ho\v rava-Witten theory was that there is a gravitational anomaly on each
ten-dimensional orbifold fixed plane, and that this can, essentially,
only be cancelled by
adding an $E_{8}$ Yang-Mills $N=1$ supermultiplet to each plane. That is, the
existence of a gravitational anomaly dictated the structure of the twisted
sector necessary to cancel it. This leads us to ask how one can determine the
twisted sector spectrum on an orbifold fixed plane, such as a seven- or
four-plane, that does not admit a gravitational anomaly. Unless this can be
done, the associated $M$-theory orbifold cannot be constructed.

Here, we explicitly answer this question. We show, within the context
of several simple but representative examples, how local anomaly cancellation
on orbifold planes with gravitational anomalies determines the twisted sector
spectrum on orbifold planes which intersect them, {\it even if these planes have no
gravitational anomalies themselves}. This result follows from the
details of anomaly cancellation which, we demonstrate, requires a chiral
spectrum {\it carrying fractional anomaly data}. Such fractional data can
only arise from a twisted sector on the intersecting fixed planes. In
addition to determining the twisted sector on orbifold fixed planes
without gravitational anomalies, {\it local anomaly freedom also uniquely specifies
additional Chern-Simons interactions that must appear on the worldvolumes} of
these planes.

In this paper, we work within the context of $M$-theory on a
background ${\bf R \rm}^{6} \times S^{1}/{\bf Z \rm}_{2} \times
T^{4}/{\bf Z \rm}_{2}$ spacetime. In this case, our results will apply to the
seven-dimensional fixed planes. This is, in some sense, the most illustrative
example possible since seven-planes can admit no chiral anomalies,
gravitational or gauge. However, it is clear that similar methods will apply
to other $M$-theory orbifolds based on threefolds, including those with
four-dimensional fixed planes and $N=1$ supersymmetry. It is expected
that one can now compute the
twisted sector spectrum on these four-planes, opening the door for new
phenomenological particle physics models. This work will be reported
on elsewhere.

\setcounter{equation}{0}
\section{$M$-Theory on $S^{1}/{\bf Z \rm}_{2} \times T^{4}/{\bf Z \rm}_{2}$
Orbifolds}

In this paper, we will, for specificity, consider $M$-theory orbifolds on
$S^{1}/{\bf Z \rm}_{2} \times T^{4}/{\bf Z \rm}_{2}$. The spacetime has topology
${\bf R \rm}^{6} \times S^{1} \times T^{4}$, where each of the
five compact coordinates takes
values on the interval $[-\pi, \pi]$ with the endpoints identified. Let
$x^{\mu}$ parameterize the six non-compact dimensions, while $x^{i}$ and
$x^{11}$ parameterize the $T^{4}$ and $S^{1}$ factors respectively. Then the
${\bf Z \rm}_{2}$ action on $S^{1}$ is defined by
\brr \alpha : (x^{\mu}, x^{i}, x^{11})
     \longrightarrow (x^{\mu}, x^{i}, -x^{11})
\label{eq:1}\err
whereas the ${\bf Z \rm}_{2}$ action on $T^{4}$ is
\brr \beta : (x^{\mu}, x^{i}, x^{11})
     \longrightarrow (x^{\mu}, -x^{i}, x^{11}) \,.
\label{eq:2}\err
The element $\alpha$ leaves invariant the two ten-planes defined by $x^{11}=0$
and $x^{11}=\pi$, while $\beta$ leaves invariant the sixteen seven-planes
defined when the four coordinates $x^{i}$ individually assume the values $0$
or $\pi$. Finally, $\alpha\beta$ leaves invariant the thirty-two six-planes
defined when all five compact coordinates individually assume the values $0$
or $\pi$. The $\alpha\beta$ six-planes coincide with the intersections of the
$\alpha$ ten-planes with the $\beta$ seven-planes. The global structure of
this orbifold is shown in Figure 1.

\begin{figure}
\begin{center}
\includegraphics[width=6in,angle=0]{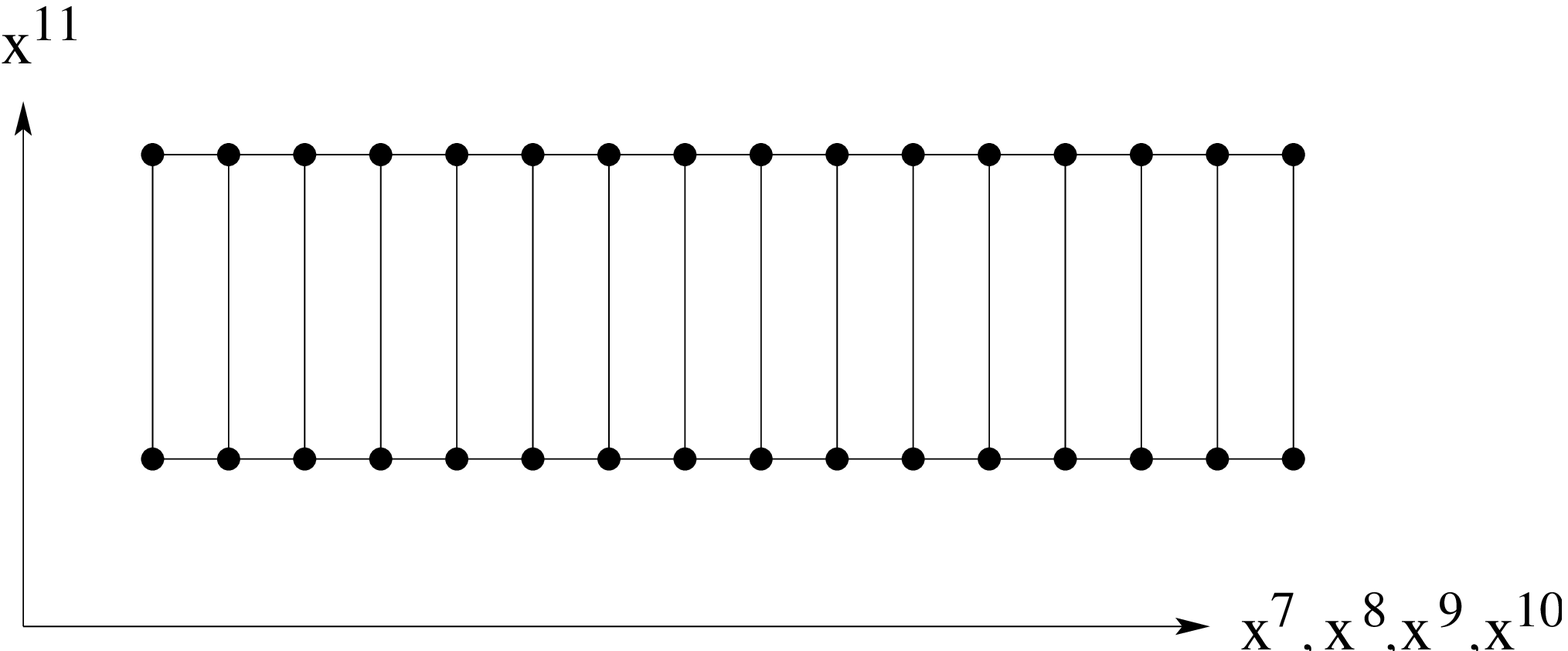}\\[.2in]
\parbox{6in}{Figure 1:  The global structure of orbifold planes
in the $S^1/{\bf Z}_2\times T^4/{\bf Z}_2$ orbifold.
The two horizontal lines represent the
two ten-dimensional (``Ho{\v r}ava-Witten") fixed planes associated
with the ${\bf Z}_2$ factor denoted $\a$, while
the sixteen vertical lines represent the seven-dimensional
fixed-planes associated with the ${\bf Z}_2$ action denoted $\b$.
The thirty-two six-dimensional fixed planes assocated with
$\a\b$ are represented by the solid dots.  These coincide
with the intersection of the $\a$ planes and the $\b$ planes. }
\end{center}
\end{figure}

In ten-dimensional superstring theories, the spectrum of the theory on an
orbifold background can be computed directly from the string equations. This
is not possible in eleven-dimensional $M$-theory, since the
fundamental underlying structure of this theory remains unknown. However, as was
demonstrated in \cite{hw1,hw2} for the case of a one-dimensional
$S^{1}/{\bf Z \rm}_{2}$
orbifold and generalized to higher dimensional orbifolds in
\cite{mlo,phase,new,ksty},
the spectrum of $M$-theory orbifolds can be determined by
exploiting the requirement that the theory be free of chiral anomalies. Since
$S^{1}/{\bf Z \rm}_{2}$ appears as a subspace of the orbifold of interest in this
paper, we begin our analysis by briefly reviewing the results of
\cite{hw1,hw2}.

\setcounter{equation}{0}
\section{The Ten-Plane Anomaly}

A gravitational anomaly arises on each ten-plane due to the coupling of chiral
projections of the bulk gravitino to currents localized on the fixed
planes. Since the two ten-planes are indistinguishable aside from their
position, this anomaly is identical on each of the two planes and can be
computed by conventional means if proper care is used. The reason why extra
care is needed is that each ten-plane anomaly arises from the coupling of
${\it eleven}$-dimensional fermions to ${\it ten}$-dimensional currents, whereas
standard index theorem results only apply to ${\it ten}$-dimensional fermions
coupled to ${\it ten}$-dimensional currents. If one notes that the index theorem
can be applied to the small radius limit where the two ten-planes
coincide, then the gravitational anomaly on each individual ten-plane can be
computed; it is simply {\it one-half} of the index theorem anomaly derived
using the ``untwisted'' sector spectrum in ten-dimensions. By untwisted sector,
we mean the ${\bf Z \rm}_{2}$ projection of the eleven-dimensional bulk space
supergravity multiplet onto each ten-dimension fixed plane. This untwisted spectrum
forms the ten-dimensional $N=1$ supergravity multiplet containing a graviton,
a chiral gravitino, a two-form and a scalar dilaton. We denote by
$R$ the ten-dimensional Riemann tensor, regarded as an $SO(9,1)$-valued form.

As pointed out in \cite{hw1,hw2}, in addition to the
untwisted spectrum, one must allow for the possibility of ``twisted'' sector
$N=1$ supermultiplets that live on each ten-dimensional orbifold plane only. For
the case at hand, the twisted sector spectrum must fall into $N=1$ Yang-Mills
supermultiplets consisting of gauge fields and chiral gauginos.
These will give rise to an additional contribution to the
gravitational anomaly on each ten-plane, as well as to mixed and
pure-gauge anomalies. However, since the twisted sector fields are
ten-dimensional, these anomalies can be computed directly from the standard formulas,
\it without multiplying by one-half \rm. The twisted sector
gauge group, the dimension of the gauge group and the gauge field strength
on the $i$-th ten-plane are denoted by ${\cal{G}}_{i}$,
$n_{i}={\rm dim}\,{\cal{G}}_{i}$
and $F_{i}$ respectively, for $i=1,2$.

The quantum mechanical one-loop local chiral anomaly on the
$i$-th ten-plane is characterized by the twelve-form
\brr I_{12}({\rm 1\!-\!loop})_{i}
     &=& \frac{1}{4}\left(I_{GRAV}^{(3/2)}(R)
     -I_{GRAV}^{(1/2)}(R)\right)
     \nonumber\\[.1in]
     & & \!\! +\frac{1}{2}\left(n_{i}\,I_{GRAV}^{(1/2)}(R)
     +I_{MIXED}^{(1/2)}(R,F_{i})+I_{GAUGE}^{(1/2)}(F_{i})\right)
\label{eq:3}\err
from which the anomaly arises by descent. The constituent polynomials contributing
to the pure gravitational anomaly due to the chiral spin $3/2$ and chiral spin
$1/2$ fermions are
\brr I_{GRAV}^{(3/2)}(R)
     =\frac{1}{(2\pi)^{5}6!}\bpl\,
     \frac{55}{56}\,{\rm tr}\,R^{6}
     -\frac{75}{128}\,{\rm tr}\,R^{4}\wedge {\rm tr}\,R^{2}
     +\frac{35}{512}\,({\rm tr}\,R^{2})^{3}\,\bpr
\label{eq:4}\err
and
\brr I_{GRAV}^{(1/2)}(R)
     =\frac{1}{(2\pi)^{5}6!}\,\bpl
     -\frac{1}{504}\,{\rm tr}\,R^{6}
     -\frac{1}{384}\,{\rm tr}\,R^{4}\wedge {\rm tr}\,R^{2}
     -\frac{5}{4608}\,(\,{\rm tr}\,R^{2})^{3}\,\bpr
\label{eq:5}\err
respectively, where ${\rm tr}$ is the trace of the $SO(9,1)$ indices.
The polynomials contributing to the mixed and pure-gauge anomalies are due to
chiral spin $1/2$ fermions only and are given by
\brr I_{MIXED}^{(1/2)}(R,F_{i})
     &=& \frac{1}{(2\pi)^{5}6!}\,\bpl\,
     \frac{1}{16}\,{\rm tr}\,R^{4}\wedge {\rm Tr}\,F_{i}^{2}
     +\frac{5}{64}\,(\,{\rm tr}\,R^{2})^{2}\wedge {\rm Tr}\,F_{i}^{2}
     \nonumber\\[.1in]
     & & \hspace{.6in}
     -\frac{5}{8}\,{\rm tr}\,R^{2}\wedge\,{\rm Tr}\,F_{i}^{4}\,\bpr
\label{eq:6}\err
and
\brr I_{GAUGE}^{(1/2)}(F_{i})
     =\frac{1}{(2\pi)^{5}6!}\,{\rm Tr}\,F_{i}^{6} \,.
\label{eq:7}\err
Here ${\rm Tr}$ is the trace over the adjoint representation of ${\cal{G}}_{i}$. All
the anomaly polynomials are computed using standard index theorems. Each term
in (\ref{eq:3}) has a factor of $1/2$ because the relevant fermions are
Majorana-Weyl with half the degrees of freedom of Weyl fermions. The
first two terms in (\ref{eq:3}) arise from untwisted sector fermions, whereas the
last three terms are contributed by the twisted sector. It follows from the
above discussion that the first two terms must have an \it
additional factor of 1/2 \rm, accounting for the overall coefficient
of $1/4$, whereas the remaining three terms are given exactly by the index theorems.

The quantum anomaly (\ref{eq:3}) would spoil the consistency of the
theory were it not to cancel against some sort of classical inflow anomaly.
Hence, it is imperative to discern the presence of appropriate
local classical counterterms to cancel against (\ref{eq:3}).
One begins the analysis of anomaly cancellation by considering
the pure ${\rm tr}\,R^{6}$ term in (\ref{eq:3}) which is irreducible
and must therefore
identically vanish. It follows from the above that this term is
\brr -\frac{1}{2(2\pi)^{5}6!}\,
     \frac{(n_{i}-248)}{494}\,{\rm tr}\,R^{6} \,.
\label{eq:8}\err
Therefore, the ${\rm tr}\,R^{6}$ term will vanish if and only if each gauge group
${\cal{G}}_{i}$ satisfies the constraint
\brr n_{i}=248 \,.
\label{eq:9}\err
Without yet specifying which $248$-dimensional gauge group is permitted, we
substitute $248$ for $n_{i}$ in (\ref{eq:3}) obtaining
\brr I_{12}({\rm 1\!-\!loop})_{i}
     &=& \frac{1}{2(2\pi)^{5}6!}\,\bpl\,
     -\frac{15}{16}\,{\rm tr}\,R^{4}\wedge {\rm tr}\,R^{2}
     -\frac{15}{64}\,(\,{\rm tr}\,R^{2})^{3}
     +\frac{1}{16}\,{\rm tr}\,R^{4}\wedge {\rm Tr}\,F_{i}^{2}
     \nonumber\\[.1in]
     & & \hspace{.7in}
     +\frac{5}{64}\,(\,{\rm tr}\,R^{2})^{2}\wedge {\rm Tr}\,F_{i}^{2}
     -\frac{5}{8}\,{\rm tr}\,R^{2}\wedge {\rm Tr}\,F_{i}^{4}
     +{\rm Tr}\,F_{i}^{6}\,\bpr \,.
\label{eq:10}\err
Although non-vanishing, this part of the anomaly is reducible. It follows that
it can be made to cancel as long as it can be factorized into the product
of two terms, a four-form and an eight-form. A necessary requirement for this
to be the case is that
\brr {\rm Tr}\,F_{i}^{6}=
     \frac{1}{24}\,{\rm Tr}\,F_{i}^{4}\wedge {\rm Tr}\,F_{i}^{2}
     -\frac{1}{3600}\,(\,{\rm Tr}\,F_{i}^{2}\,)^{3} \,.
\label{eq:11}\err
There are two Lie groups with dimension $248$
that satisfy this condition, the non-Abelian group
$E_{8}$ and the Abelian group $U(1)^{248}$. Both groups represent allowed
twisted matter gauge groups on each ten-plane. Hence, from anomaly
considerations alone one can determine the twisted sector on each ten-plane,
albeit with a small ${\it ambiguity}$ in the allowed
twisted sector gauge group. In this paper, we consider only the non-Abelian
gauge group $E_{8}$. Using (\ref{eq:11}) and several $E_{8}$ trace relations,
the anomaly polynomial (\ref{eq:10}) can be re-expressed as follows
\brr I_{12}({\rm 1\!-\!loop})_{i}=
     \frac{1}{3}\,\pi\,I_{4(i)}^{3}
     + X_{8} \wedge I_{4(i)}
\label{eq:12}\err
where $X_{8}$ is the eight-form
\brr X_{8}=
     \frac{1}{(2\pi)^{3}4!}\,\bpl\,
     \frac{1}{8}\,{\rm tr}\,R^{4}
     -\frac{1}{32}\,(\,{\rm tr}\,R^{2}\,)^{2}\,\bpr
\label{eq:13}\err
and $I_{4\,(i)}$ is the four-form given by
\brr I_{4\,(i)}=
    \frac{1}{16\pi^{2}}\,\bpl\,
    \frac{1}{30}\,{\rm Tr}\,F_{i}^{2}
    -\frac{1}{2}\,{\rm tr}\,R^{2}\bpr \,.
\label{eq:14}\err
Once in this factorized form, the anomaly $I_{12}({\rm 1\!-\!loop})_{i}$
can be cancelled as follows.

First, the Bianchi identity $dG=0$, where G is the field strength of the
three-form $C$ in the eleven-dimensional supergravity multiplet, is modified to
\brr dG=\sum_{i=1}^{2}I_{4(i)} \wedge \delta_{M_{i}^{10}}^{(1)}
\label{eq:15}\err
where $I_{4(i)}$ is the four-form given in (\ref{eq:14}) and
$\delta_{M_{i}^{10}}^{(1)}$ is a one-form brane current with support on the
$i$-th ten-plane. Second, we note that the eleven-dimensional supergravity
action contains the terms
\brr S=\cdots -\frac{\pi}{3}
     \int C \wedge G \wedge G+ \int G \wedge X_{7}
\label{eq:16}\err
where $X_{7}$ satisfies $dX_{7}=X_{8}$.
The $CGG$ interaction is required by the minimally-coupled supergravity
action, while the $GX_{7}$ term  is an additional higher-derivative
interaction necessitated by five-brane anomaly cancellation. Using the
modified Bianchi identity (\ref{eq:15}), one can compute the variation of
these two terms under Lorentz and gauge transformations. The result is
that the $CGG$ and $GX_{7}$ terms have classical anomalies which descend from
the polynomials
\brr I_{12}(CGG)_{i}=-\frac{\pi}{3}I_{4\,(i)}^{\,3}
\label{eq:17}\err
and
\brr I_{12}(GX_{7})_{i}=- X_{8} \wedge I_{4(i)} \,.
\label{eq:18}\err
respectively. It follows that
\brr I_{12}({\rm 1\!-\!loop})_{i}
     +I_{12}(CGG)_{i}
     +I_{12}(GX_{7})_{i}=0
\label{eq:19}\err
and, hence, the total anomaly cancels exactly.

We conclude that the requirement of local anomaly
cancellation on the each of the two $S^{1}/{\bf Z \rm}_{2}$ orbifold
ten-planes specifies the twisted spectrum of the theory. This specification is
almost, but not quite, unique, allowing $N=1$ vector supermultiplets
with either gauge group $E_{8}$ or $U(1)^{248}$. An important ingredient in
this analysis was the fact that the contribution to the anomaly on each
ten-plane from the untwisted sector was {\it a factor of 1/2}  smaller than the
index theorem result. This followed from the fact that the index theorem
had to be spread over two equivalent ten-planes. A direct consequence of
this is that the non-Abelian gauge group on each ten-plane is $E_{8}$, not
$E_{8} \times E_{8}$, and that the gauge group $SO(32)$
is disallowed. Since
$S^{1}/{\bf Z \rm}_{2}$ is a subspace of $S^{1}/{\bf Z \rm}_{2}
\times T^{4}/{\bf Z \rm}_{2}$, the results of this section continue to hold on
the larger orbifold. We now discuss the cancellation of local anomalies in the
other factor space, $T^{4}/{\bf Z \rm}_{2}$.

\setcounter{equation}{0}
\section{The Seven-Plane Anomaly}

The quantum anomalies on each of the sixteen indistinguishable
seven-planes of the $T^{4}/{\bf Z \rm}_{2}$ orbifold are easy to
analyze. In analogy with the ten-planes, an untwisted sector is induced on
each seven-plane by the ${\bf Z \rm}_{2}$ projection of the
eleven-dimensional supergravity  multiplet. This untwisted spectrum forms the
seven-dimensional $N=1$ supergravity multiplet consisting of a graviton,
a gravitino, three vector fields, a two-form, a real scalar dilaton
and a spin 1/2 dilitino.
However, unlike the case of a ten-plane, gravitational anomalies
cannot be supported on a seven-plane. In fact, since there are no chiral fermions in
seven-dimensions, no chiral anomaly of any kind, gravitational or gauge, can arise.
Hence, with no local chiral anomalies to cancel, it would appear to be
impossible to compute the twisted sector spectrum of any seven-plane. As long
as we focus on the seven-planes exclusively, this conclusion is correct. However, as
we will see in the next section, {\it the cancellation of the local anomalies on
the thirty-two six-dimensional $\alpha\beta$ orbifold planes, formed from the
intersection of the $\alpha$ ten-planes with the $\beta$ seven-planes, will
require a non-vanishing twisted sector spectrum on each
seven-plane and dictate its structure}. With this in mind, we now turn to
the analysis of anomalies localized on the intersection six-planes in the
full $S^{1}/{\bf Z \rm}_{2} \times T^{4}/{\bf Z \rm}_{2}$ orbifold.

\setcounter{equation}{0}
\section{The Six-Plane Anomaly and Seven-Plane Twisted Sector}

As in the case for the ten-planes, a gravitational anomaly will arise on each
six-plane due to the coupling of chiral projections of the bulk gravitino to
currents localized on the thirty-two fixed planes. Since the thirty-two
six-planes are indistinguishable, the anomaly is the same on each
plane and can be computed by conventional means if proper care is
taken. Noting that the standard index theorems can be applied to the small
radius limit where the thirty-two six-planes coincide, it follows that the
gravitational anomaly on each six-plane is simply {\it one-thirty-second} of
the index theorem anomaly derived using the untwisted sector spectrum in
six-dimensions. In this case, the untwisted sector spectrum  is the
${\bf Z \rm}_{2} \times {\bf Z \rm}_{2}$ projection of the
eleven-dimensional bulk supergravity
multiplet onto each six-dimensional fixed plane. This untwisted spectrum forms
several $N=1$ six-dimensional supermultiplets. Namely,
the supergravity multiplet consisting
of a graviton, a chiral gravitino and a self-dual two-form,
four hypermultiplets each with four
scalars and an anti-chiral hyperino,
and one tensor multiplet with one anti-self-dual
two-form, one scalar and an anti-chiral spin 1/2 fermion.
A one-loop quantum
gravitational anomaly then arises from one chiral spin $3/2$ fermion, five
anti-chiral spin $1/2$ fermions and one each of self-dual and anti-self-dual
tensors. However, the anomalies due to the tensors cancel each other. Noting
that a chiral anomaly in six-dimensions is characterized by an eight-form,
from which the anomaly arises by descent, we find, for the $i$-th six-plane,
that
\brr I_{8}(SG)_{i}=
     \frac{1}{32}\,\bpl\,
     I_{GRAV}^{(3/2)}(R)
     -5\,I_{GRAV}^{(1/2)}(R)\,\bpr
\label{eq:20}\err
where
\brr I_{GRAV}^{(3/2)}(R)=
     \frac{1}{(2\pi)^{3}4!}\,\bpl
     -\frac{49}{48}\,{\rm tr}\,R^{4}
     +\frac{43}{192}\,(\,{\rm tr}\,R^{2}\,)^{2}\,\bpr
\label{eq:21}\err
and
\brr I_{GRAV}^{(1/2)}(R)=
    \frac{1}{(2\pi)^{3}4!}\,\bpl
    -\frac{1}{240}\,{\rm tr}\,R^{4}
    -\frac{1}{192}\,(\,{\rm tr}\,R^{2})^{2}\,\bpr \,,
    \label{eq:22}\err
where $R$ is the six-dimensional Riemann tensor, regarded as an $SO(5,1)$-valued
form. Note that the terms in brackets in (\ref{eq:20}) are the anomaly as
computed by the index theorem.
$I_{8}(SG)_{i}$ is obtained from that result by dividing by $32$.

Noting that each six-plane is embedded in one of the two ten-dimensional
planes, we see that there are additional ``untwisted'' sector fields on each
six-plane. These arise from the $\beta$
${\bf Z \rm}_{2}$ projection of the $N=1$ $E_{8}$ Yang-Mills supermultiplet on
the associated ten-plane. Such fields are untwisted from the point of view of
the six-dimensional plane, although they arise from fields that were in the
twisted sector of the ten-plane. In this paper, we will assume
that the $\beta$ action on the ten-dimensional vector multiplets does not break
the $E_{8}$ gauge group. A discussion of the case where $E_{8}$ is broken to
a subgroup by the action of $\beta$ can be found in \cite{phase,new}.
A ten-dimensional
$N=1$ vector supermultiplet decomposes in six-dimensions into an $N=1$ vector
multiplet and an $N=1$ hypermultiplet. However, the action of $\beta$ projects
out the hypermultiplet. Therefore, the ten-plane contribution to the untwisted
sector of each six-plane is an $N=1$ $E_{8}$ vector supermultiplet, which
consists of gauge fields and chiral gauginos. The gauginos contribute to the
gravitational anomaly on each six-plane, as well as adding mixed and
$E_{8}$ gauge anomalies. Noting that the standard index theorems can be
applied to the small radius limit, where each ten-plane shrinks to zero size
and, hence,  the sixteen six-planes it contains coincide,
it follows that the anomaly is
simply {\it one-sixteenth} of the index theorem result. We find that the
one-loop quantum contribution of this $E_{8}$ supermultiplet to the
gravitational, mixed and $E_{8}$ gauge anomalies on the $i$-th six-plane is
\brr I_{8}(E_{8})_{i}=
     \frac{1}{16}\,\bpl\,248\,I_{GRAV}^{(1/2)(R)}
     +I_{MIXED}^{(1/2)}(R,F_{i})
     +I_{GAUGE}^{(1/2)}(F_{i})\,\bpr
\label{eq:23}\err
where
\brr I_{MIXED}^{(1/2)}(R,F_{i})=
     \frac{1}{(2\pi)^{3}4!}\,\bpl\,
     \frac{1}{4}\,{\rm tr}\,R^{2}\wedge {\rm Tr}\,F_{i}^{2}\,\bpr
\label{eq:24}\err
and
\brr I_{GAUGE}^{(1/2)}(F_{i})=
     \frac{1}{(2\pi)^{3}4!}\,\bpl
     -{\rm Tr}\,F_{i}^{4}\,\bpr \,.
\label{eq:25}\err
Here ${\rm Tr}$ is over the adjoint ${\bf 248 \rm}$ representation of $E_{8}$.
Note that the terms in brackets in (\ref{eq:23}) are the index theorem
anomaly. $I_{8}(E_{8})_{i}$ is obtained from that result by dividing by $16$.

Are there other sources of untwisted sector anomalies on a six-plane? The
answer is, potentially yes. We note that, in addition to being embedded in one
of the two ten-planes, each six-plane is also embedded in one of the sixteen
seven-dimensional orbifold planes. In analogy with the discussion above, {\it if
there were to be a non-vanishing twisted sector spectrum on each seven-plane,
then this could descend under the $\alpha$ ${\bf Z \rm}_{2}$ projection as an
addition to the untwisted spectrum on each six-plane. This additional
untwisted spectrum could then contribute to the chiral anomalies on the
six-plane}. However, as  noted above, {\it a priori},
there is no reason for one
to believe that there is any twisted sector on a seven-dimensional orbifold
plane. Therefore, for the time being, let us assume that there is no such
contribution to the six-dimensional anomaly. We will see below that this
assumption must be carefully revisited.

As for the ten-dimensional planes, one must allow for the
possibility of twisted sector $N=1$ supermultiplets on each of the
thirty-two six-planes. The most general allowed spectrum on the $i$-th
six-plane  would be $n_{Vi}$ vector multiplets
transforming in the adjoint
representation of some as yet unspecified gauge group ${\cal{G}}_{i}$, $n_{Hi}$
hypermultiplets transforming under some representation (possibly reducible)
${\cal{R}}$ of ${\cal{G}}_{i}$, and $n_{Ti}$ gauge-singlet tensor multiplets.
We denote by ${\cal{F}}_{i}$ the gauge field strength.
Since these fields are in the twisted sector, their contribution to the chiral
anomalies can be determined directly from the index theorems without
modification. We find that the one-loop quantum contribution of the twisted
spectrum to the gravitational, mixed and ${\cal{G}}_{i}$ gauge anomalies on
the $i$-th six-plane is
\brr I_{8}({\cal{G}}_{i})
     &=& (n_{V}-n_{H}-n_{T})_{i}\,I_{GRAV}^{(1/2)}(R)
     -n_{Ti}\,I_{GRAV}^{({\rm 3\!-\!form})}(R)
     \nonumber\\[.1in]
     & & +I_{MIXED}^{(1/2)}(R,{\cal{F}}_{i})
     +I_{GAUGE}^{(1/2)}({\cal{F}}_{i})
\label{eq:26}\err
where $I_{GRAV}^{(1/2)}(R)$ is given in (\ref{eq:22}) and
\brr I_{GRAV}^{({\rm 3\!-\!form})}(R)=
    \frac{1}{(2\pi)^{3}4!}\,\bpl
    -\frac{7}{60}\,{\rm tr}\,R^{4}
    +\frac{1}{24}\,(\,{\rm tr}\,R^{2}\,)^{2}\,\bpr \,.
\label{eq:27}\err
Furthermore, the mixed and pure-gauge anomaly polynomials are modified to
\brr I_{MIXED}^{(1/2)}(R,{\cal{F}}_{i})=
     \frac{1}{(2\pi)^{3}4!}\,\bpl\,
     \frac{1}{4}\,{\rm tr}\,R^{2}\wedge
     {\rm trace}\,{\cal{F}}_{i}^{2}\,\bpr
\label{eq:28}\err
and
\brr I_{GAUGE}^{(1/2)}({\cal{F}}_{i})=
     \frac{1}{(2\pi)^{3}4!}\,\bpl
     -{\rm trace}\,{\cal{F}}_{i}^{4}\,\bpr \,,
\label{eq:29}\err
where
\brr {\rm trace}\,{\cal{F}}_{i}^{n}=
     {\rm Tr}\,{\cal{F}}_{i}^{n}
     -\sum_{\alpha}\,h_{\alpha}\,{\rm tr}_{\alpha}{\cal{F}}_{i}^{n} \,.
\label{eq:30}\err
Here ${\rm Tr}$ is an adjoint trace, $h_{\alpha}$ is the number of hypermultiplets
transforming in the ${\cal{R}}_{\alpha}$ representation and $tr_{\alpha}$ is a
trace over the ${\cal{R}}_{\alpha}$ representation. Note that the total number
of vector multiplets is $n_{Vi}={\rm dim}\,({\cal{G}}_{i})$,
while the total number of
hypermultiplets is
$n_{Hi}=\sum_{\alpha}h_{\alpha} \times {\rm dim}\,({\cal{R}}_{\alpha})$.
The relative minus sign in (\ref{eq:30}) reflects
the anti-chirality of the hyperinos.

Combining the contributions from the two untwisted sector sources and the
twisted sector, the total one-loop quantum anomaly on the $i$-th six-plane is
the sum
\brr I_{8}({\rm 1\!-\!loop})_{i}=
     I_{8}(SG)_{i}+I_{8}(E_{8})_{i}
     +I_{8}({\cal{G}}_{i})
\label{eq:31}\err
where $I_{8}(SG)_{i}$, $I_{8}(E_{8})_{i}$ and $I_{8}({\cal{G}}_{i})$ are given
in (\ref{eq:20}), (\ref{eq:23}) and (\ref{eq:26}) respectively.

Unlike the case for the ten-dimensional planes, the classical anomaly
associated with the $GX_{7}$ term in the eleven-dimensional action
(\ref{eq:16}) can contribute to the irreducible curvature term which, in
six-dimensions, is ${\rm tr}\,R^{4}$. Therefore, our
next step is to further modify the Bianchi identity for $G=dC$ from expression
(\ref{eq:15}) to
\brr dG=\sum_{i=1}^{2}I_{4(i)} \wedge \delta_{M_{i}^{10}}^{(1)}
     +\sum_{i=1}^{32}g_{i}\,\delta_{M^{6}_{i}}^{(5)}
\label{eq:32}\err
where $\delta_{M^{6}_{i}}^{(5)}$ has support on the six-planes
$M^{6}_{i}$. As discussed in \cite{mlo,phase}, the magnetic charges
$g_{i}$ are required to take the values
\brr g_{i}=-3/4, -1/4, +1/4,...
\label{eq:33}\err
Using the modified Bianchi identity (\ref{eq:31}), one can compute the
variation of the $GX_{7}$ term under Lorentz  and gauge
transformations. The result is that this term gives rise to a classical
anomaly that descends from the polynomial
\brr I_{8}(GX_{7})_{i}=-g_{i}\,X_{8}
\label{eq:35}\err
where $X_{8}$ is presented in expression (\ref{eq:13}).
The relevant anomaly is then
\brr I_{8}({\rm 1\!-\!loop})_{i}
     +I_{8}(GX_{7})_{i}
\label{eq:36}\err
where $I_{8}({\rm 1\!-\!loop})_{i}$ is given in (\ref{eq:31}).
This anomaly spoils the consistency of the theory and, hence, must cancel. One
begins the analysis of anomaly cancellation by considering the pure
${\rm tr}\,R^{4}$
term in (\ref{eq:36}) which is irreducible and must identically vanish.
It follows from the above that this term is
\brr -\frac{1}{(2\pi)^{3}4!\,240}\,
    (n_{Vi}-n_{Hi}-29n_{Ti}
    +30\,g_{i}+23)\,{\rm tr}\,R^{4} \,.
\label{eq:37}\err
Therefore, the ${\rm tr}\,R^{4}$ term will vanish if and only if on each orbifold
plane the constraint
\brr n_{Vi}-n_{Hi}-29n_{Ti}+30g_{i}+23=0
\label{eq:38}\err
is satisfied. {\it Herein lies a problem, and the main point of this paper}.
Noting from (\ref{eq:33}) that $g_{i}=c_{i}/4$ where $c_{i}=-3,-1,1,3,5,...$,
we see that cancelling the ${\rm tr}\,R^{4}$ term requires that we satisfy
\brr n_{Vi}-n_{Hi}-29n_{Ti}=(-15c_{i}-46)/2 \,.
\label{eq:39}\err
However, {\it this is not possible since the left hand side of this expression
is an integer and the right hand side always half integer}. There is only one
possible resolution of this problem, which is to carefully review the
only assumption that was made above, that is, that
there is no twisted sector on a seven-plane and,
hence, no contribution of the seven-planes by $\alpha$ ${\bf Z \rm}_{2}$
projection to the untwisted anomaly on a six-plane. As we now show,
{\it this assumption is false}.

Let us now allow for the possibility that there is a twisted sector of $N=1$
supermultiplets on each of the sixteen seven-planes. The most general allowed
spectrum on the $i$-th seven-plane would be $n_{7Vi}$
vector supermultiplets transforming in the adjoint representation of some as yet
unspecified gauge group $G_{7i}$. Each seven-dimensional vector
multiplet contains a gauge field, three scalars and a gaugino. With respect to
six-dimensions, this vector multiplet decomposes into an $N=1$ vector
supermultiplet and a single hypermultiplet. Under the
$\alpha$ ${\bf Z \rm}_{2}$ projection to each of the two embedded six-planes,
the gauge group $G_{7i}$ can be preserved or broken to a subgroup. In
either case, we denote the six-dimensional gauge group arising in this manner
as $\tilde{\cal{G}}_{i}$, define $\tilde{n}_{Vi}={\rm dim}\,\tilde{\cal{G}}_{i}$ and
write the associated gauge field strength as $\tilde{\cal{F}}_{i}$.
In this paper, for simplicity, we will assume that the
gauge group is unbroken by the orbifold projection, that is,
$\tilde{\cal{G}}_{i}=G_{7i}$. The more general case where it is broken to a
subgroup is discussed in \cite{phase,new}. Furthermore, the $\alpha$ action
projects out either the six-dimensional vector supermultiplet, in which case
the hypermultiplet descends to the six-dimensional untwisted sector,
or the six-dimensional hypermultiplet, in which case the vector supermultiplet
enters the six-dimensional untwisted sector. We denote by $\tilde{n}_{Hi}$ the
number of hypermultiplets arising in the six-dimensional untwisted sector
by projection from the seven-plane, and specify their (possibly reducible)
representation
under $\tilde{\cal{G}}_{i}$ as $\tilde{\cal{R}}$. Since these fields are in
the untwisted sector associated with a single seven-plane, and since there are
two six-planes embedded in each seven-plane, their contribution to the quantum
anomaly on each six-plane can be determined by taking {\it 1/2 of the index
theorem result}. We find that the one-loop quantum contribution of this part of
the the untwisted spectrum to the gravitational, mixed and $\tilde{\cal{G}}_{i}$
gauge anomalies on the $i$-th six-plane is
\brr I_{8}(\tilde{\cal{G}}_{i})=
     \frac{1}{2}\,\bpl\,
     (\tilde{n}_{V}-\tilde{n}_{H})_{i}\,I_{GRAV}^{(1/2)}(R)
     +I_{MIXED}^{(1/2)}(R,\tilde{{\cal{F}}}_{i})
     +I_{GAUGE}^{(1/2)}(\tilde{{\cal{F}}}_{i})\,\bpr
\label{eq:40}\err
where $I_{GRAV}^{(1/2)}(R), I_{MIXED}^{(1/2)}(R,\tilde{{\cal{F}}}_{i})$
and $I_{GAUGE}^{(1/2)}(\tilde{{\cal{F}}}_{i})$
are given in (\ref{eq:22}),(\ref{eq:28}) and (\ref{eq:29}) respectively with
the gauge and hypermultiplet quantities replaced by their `` $\sim$ ''
equivalents.

The total quantum anomaly on the $i$-th six-plane is now modified to
\brr I_{8}({\rm 1\!-\!loop})_{i}
     +I_{8}(\tilde{\cal{G}}_{i})
\label{eq:41}\err
where $I_{8}({\rm 1\!-\!loop})_{i}$ and $I_{8}(\tilde{\cal{G}}_{i})$ are given
in (\ref{eq:31}) and (\ref{eq:40}) respectively. It follows that the relevant
anomaly contributing to, among other things, the irreducible ${\rm tr}\,R^{4}$ term
is modified to
\brr I_{8}({\rm 1\!-\!loop})_{i}
     +I_{8}(\tilde{\cal{G}}_{i})
     +I_{8}(GX_{7})_{i} \,.
\label{eq:42}\err
This anomaly spoils the quantum consistency of the theory and, hence, must
cancel. We again begin by considering the pure ${\rm tr}\,R^{4}$ term in
(\ref{eq:42}).
This term is irreducible and must identically vanish. It follows from the
above that this term is
\brr -\frac{1}{(2\pi)^{3}4!\,240}\,
    (\,n_{Vi}-n_{Hi}
    +\ft12\,\tilde{n}_{Vi}
    -\ft12\,\tilde{n}_{Hi}
    -29n_{Ti}+30g_{i}+23\,)\,{\rm tr}\,R^{4} \,.
\label{eq:43}\err
Therefore, the ${\rm tr}\,R^{4}$ term will vanish if and only if on each orbifold
plane the constraint
\brr n_{Vi}-n_{Hi}
    +\ft12\,\tilde{n}_{Vi}-\ft12\,\tilde{n}_{Hi}
    -29\,n_{Ti}+30\,g_{i}+23=0
\label{eq:44}\err
is satisfied. Again, noting that $g_{i}=c_{i}/4$ where $c_{i}=-3,-1,1,3,5,...$,
we see that we must satisfy
\brr n_{Vi}-n_{Hi}
     +\ft12\,\tilde{n}_{Vi}-\ft12\,\tilde{n}_{Hi}
     -29\,n_{Ti}=\ft12\,(\,-15c_{i}-46\,) \,.
\label{eq:45}\err
As above, the right hand side is always a half integer. Now, however,
{\it because of the addition of the untwisted spectrum arising from the
seven-plane, the left hand side can also be chosen to be half integer. Hence,
the pure ${\rm tr}\,R^{4}$ term can be cancelled}.

Having cancelled the irreducible ${\rm tr}\,R^{4}$ term, we now compute the
remaining terms in the anomaly eight-form. In addition to the contributions
from (\ref{eq:42}), we must also take into account the classical anomaly
associated with the $CGG$ term in the eleven-dimensional action (\ref{eq:16}).
Using the modified Bianchi identity (\ref{eq:31}), one can compute the
variation of the $CGG$ term under Lorentz and gauge transformations. The
result is that this term gives rise to a classical anomaly that descends from
the polynomial
\brr I_{8}(CGG)_{i}=
     -\pi\,g_{i}\,I_{4\,(i)}^{\,2}
\label{eq:46}\err
where $I_{4\,(i)}$ is given in expression (\ref{eq:14}). Adding this anomaly to
(\ref{eq:42}), and cancelling the ${\rm tr}\,R^{4}$ term by imposing constraint
(\ref{eq:44}), we can now determine the remaining terms in the anomaly
eight-form.

Recall that, in this paper, we are assuming that the $\beta$ action on the
ten-dimensional vector supermultiplet does not break the $E_{8}$ gauge group.
In this case, we can readily show that there can be no twisted sector vector
multiplets on any six-plane. Rather than complicate the present discussion, we
will simply assume here that gauge field strengths ${\cal{F}}_{i}$ do not
appear. Furthermore, cancellation of the
complete anomaly, in the case where $E_{8}$ is unbroken, requires that
$\tilde{{\cal{G}}}_{i}$ be a product of $U(1)$ factors. Here, we will limit
the discussion to the simplest case where
\brr \tilde{{\cal{G}}}_{i}=U(1)
\label{eq:a}\err
The $\beta$ action on the seven-dimensional plane then
either projects a single vector supermultiplet, or a single
chargeless hypermultiplet,
onto the untwisted sector of the six-plane. In either case, no $U(1)$ anomaly
exists. Hence, the gauge field strengths $\tilde{{\cal{F}}}_{i}$ also do not
appear. With this in mind, we now compute the remaining terms in the anomaly
eight-form. They are
\brr & & \frac{1}{(2\pi)^{3}4!\,16}\,\bpl\,
     \ft34\,(\,1-4\,n_{Ti}\,)\,({\rm tr}\,R^{2})^{2}
     +\ft{1}{20}\,(\,5+8\,g_{i}\,)\,
     {\rm tr}\,R^{2}\wedge{\rm Tr}\,F_{i}^{2}
     \nonumber\\[.1in]
     & & \hspace{.8in}
     -\ft{1}{100}\,(\,1+\ft43\,g_{i}\,)\,
     (\,{\rm Tr}\,F_{i}^{2})^{2}\,\bpr
\label{eq:47}\err
where we have used the $E_{8}$ trace relation
${\rm Tr}\,F^{4}=\frac{1}{100}({\rm Tr}\,F^{2})^{2}$.
Note that, since $n_{Ti}$ is a non-negative integer and $g_{i}$ must satisfy
(\ref{eq:33}), the first two terms of this expression term can never vanish.
Furthermore, it is straightforward to show that (\ref{eq:47}) will
factor into an exact square, and, hence, be potentially cancelled by a
six-plane Green-Schwarz mechanism, if and only if
\brr 4\,(\,4\,n_{Ti}-1\,)(\,3+4\,g_{i}\,)=(\,5+8\,g_{i}\,)^{2}
\label{eq:48}\err
Again, this equation has no solutions for the allowed values of $n_{Ti}$ and
$g_{i}$. It follows that anomaly (\ref{eq:47}), as it presently stands, cannot
be be made to identically vanish or cancel. The resolution of this problem
was first described in \cite{mlo}, and
consists of the realization that {\it the existence of seven-planes in
the theory necessitates the introduction of additional Chern-Simons
interactions in the action}, one for each seven-plane. The required terms are
\brr S=\cdots
     +\sum_{i=1}^{16}\int
     \delta_{M_{i}^{7}}^{(4)}\wedge G \wedge Y_{3(i)}^{0}
\label{eq:49}\err
where $dY_{3(i)}^{0}=Y_{4(i)}$ is a gauge-invariant four-form polynomial.
$Y_{4(i)}$ arises from the curvature $R$ and also the field strength
$\tilde{\cal{F}}_{i}$ associated with the ${\it additional}$ adjoint
super-gauge fields localized on the $i$-th seven-plane. It is given by
\brr Y_{4(i)}=
     \frac{1}{4\pi}\bpl
     -\frac{1}{32}\,\eta\,{\rm tr}\,R^{2}
     +\rho\,{\rm tr}\,\tilde{\cal{F}}_{i}\,\bpr
\label{eq:50}\err
where $\eta$ and $\rho$ are rational coefficients. Using the modified Bianchi
identity (\ref{eq:32}), one can compute the variation of the
$\delta^{7}GY_{3}$ terms under Lorentz and gauge transformations. The
result is that these give rise to a classical anomaly that descends from the
polynomial
\brr I_{8}(\delta^{7}GY_{3})_{i}=
     -I_{4\,(i)} \wedge Y_{4(i)}
\label{eq:51}\err
where $I_{4\,(i)}$ is the four-form given in (\ref{eq:14}).

The total anomaly on the $i$-th six-plane is now modified to
\brr I_{8}({\rm 1\!-\!loop})_{i}
     +I_{8}(\tilde{\cal{G}}_{i})
     +I_{8}(GX_{7})_{i}
     +I_{8}(CGG)_{i}
     +I_{8}(\delta^{7}GY_{3})_{i}
\label{eq:52}\err
where $I_{8}({\rm 1\!-\!loop})_{i}, I_{8}(\tilde{\cal{G}}_{i}),
I_{8}(GX_{7})_{i}, I_{8}(CGG)_{i}$ and $I_{8}(\delta^{7}GY_{3})_{i}$
are given in (\ref{eq:31}), (\ref{eq:40}), (\ref{eq:35}), (\ref{eq:46}) and
(\ref{eq:51}) respectively. Note that for the fixed plane intersection
presently under discussion, the field strength $\tilde{\cal{F}}_{i}$ does
not enter the anomaly eight-form (\ref{eq:47}). Therefore, within this context,
we must take
\brr \rho=0 \,.
\label{eq:add}\err
We will exhibit an example of non-vanishing $\rho$ parameter at the end of
this section. After cancelling the irreducible ${\rm tr}\,R^{4}$ term,
the remaining anomaly now becomes
\brr & & \frac{1}{(2\pi)^{3}4!16}\bpl\,
     \frac{3}{4}\,(1-4n_{Ti}-\eta)\,({\rm tr}\,R^{2})^{2}
     \nonumber\\[.1in]
     & & \hspace{.8in}
     +\frac{1}{20}\,(5+8g_{i}+\eta)\,{\rm tr}\,R^{2}\wedge {\rm Tr}\,F_{i}^{2}
     -\frac{1}{100}\,(1+\frac{4}{3}g_{i})\,({\rm Tr}\,F_{i}^{2})^{2}\,\bpr
\label{eq:53}\err
Depending on the number of untwisted hypermultiplets, $n_{Ti}$, these terms
{\it can} be made to cancel or to factor into the sum of exact squares. In
this paper, we consider the $n_{Ti}=0,1$ cases only. As discussed in
\cite{phase,new}, the solutions where $n_{Ti}\geq2$ are related to the
$n_{Ti}=0,1$ solutions by the absorption of one or more five-branes
from the bulk space onto the $i$-th six-plane.

We first consider the case where
\brr n_{Ti}=0 \,.
\label{eq:b}\err
In this case, no further Green-Schwarz type mechanism
in six-dimensions is possible and the anomaly must vanish identically. We see
from (\ref{eq:53}) that this is possible if and only if
\brr g_{i}=-3/4, \qquad \eta=1 \,.
\label{eq:54}\err
{\it It is important to note that this solution only exists
for a non-vanishing value of parameter $\eta$. Hence, the additional
Chern-Simons interactions (\ref{eq:49}) are essential for the anomaly
to vanish identically in the $n_{Ti}=0$ case}.
Inserting these results into expression (\ref{eq:44}) for the vanishing of the
irreducible ${\rm tr}\,R^{4}$ term, and recalling that $n_{Vi}=0$, we find that
\brr -2n_{Hi}+\tilde{n}_{Vi}-\tilde{n}_{Hi}=-1 \,.
\label{eq:55}\err
Equation (\ref{eq:55}) can be solved in several ways.
Remembering that $\tilde{{\cal{G}}}_{i}=U(1)$,
the first solution then consists of allowing the $U(1)$ hypermultiplet to
descend to the six-plane while projecting out the $U(1)$ vector multiplet.
Equation (\ref{eq:55}) is then solved by taking the number of twisted
hypermultiplets to vanish. That is, take
\brr \tilde{n}_{Hi}=1, \qquad \tilde{n}_{Vi}=0, \qquad n_{Hi}=0 \,.
\label{eq:56}\err
The second solution follows by doing the reverse, that is, projecting out the
$U(1)$ hypermultiplet and allowing the $U(1)$ vector multiplet to descend to
the six-plane. In this case, equation (\ref{eq:55}) is solved by taking
\brr \tilde{n}_{Hi}=0, \qquad \tilde{n}_{Vi}=1, \qquad n_{Hi}=1 \,.
\label{eq:57}\err
These two solutions are illustrated in Figure 2 (a) and (b) respectively.
\begin{figure}
\begin{center}
\includegraphics[width=4in,angle=0]{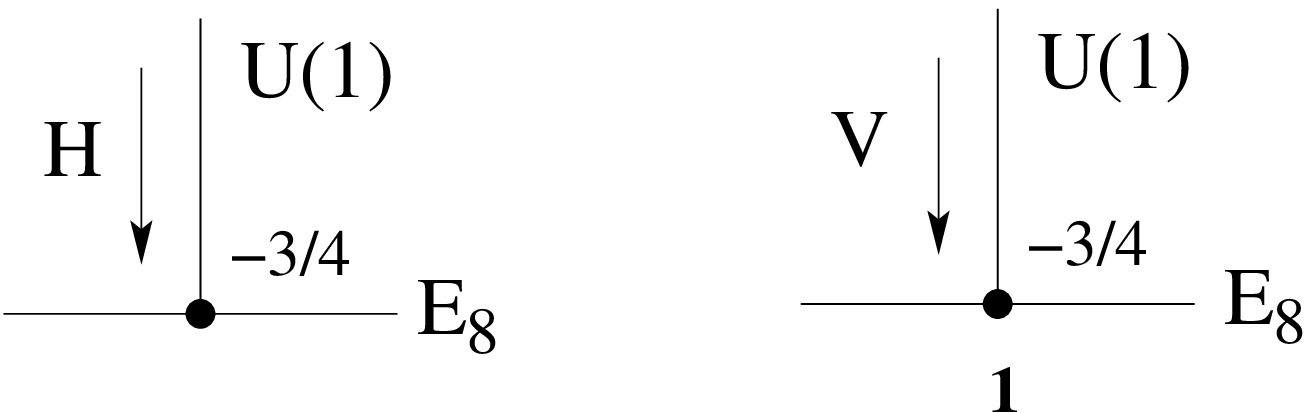}\\[.2in]
\parbox{6in}{Figures 2a and 2b: The two solutions with unbroken
ten-dimensional $E_8$
and $n_T=0$ in which the seven-dimensional gauge group is $U(1)$.
The left-hand figure depicts the case where only a chargeless
hypermultiplet survives the ${\bf Z}_2$ projection.  The right-hand figure
depicts the case where only a $U(1)$ vector multiplet survives. In the second
case we also require an extra six-dimensional singlet hypermultiplet, which is
depicted by the ${\bf 1}$ sitting below the vetex. In  both cases,
we require $\eta$=1, $\rho=0$ and $g=-3/4$.}
\end{center}
\end{figure}

Let us now consider the case where
\brr n_{Ti}=1 \,.
\label{eq:58}\err
In this case, the anomaly
(\ref{eq:53}) can be removed by a six-dimensional Green-Schwarz mechanism
as long as it factors into an exact square. It is straightforward to show that
this will be the case if and only if
\brr 4\,(\,3+\eta\,)\,(\,3+4\,g_{i}\,)=(\,5+8\,g_{i}+\eta\,)^{2} \,.
\label{eq:59}\err
This equation has two solutions
\brr g_{i}=-3/4, \qquad \eta=1
\label{eq:60}\err
and
\brr g_{i}=1/4, \qquad \eta=1 \,.
\label{eq:61}\err
{\it Again, note that these solutions require a non-vanishing value of the
parameter $\eta$. Hence, the additional Chern-Simons interactions
(\ref{eq:49}) are also essential for anomaly factorization in the $n_{Ti}=1$
case}.
Inserting these into the expression for the vanishing of the irreducible
${\rm tr}\,R^{4}$ term, and recalling that $n_{Vi}=0$, we find
\brr -2n_{Hi}+\tilde{n}_{Vi}-\tilde{n}_{Hi}=57
\label{eq:62}\err
and
\brr -2n_{Hi}+\tilde{n}_{Vi}-\tilde{n}_{Hi}=-3 \,.
\label{eq:63}\err
The first equation (\ref{eq:62}) cannot be solved within the context of
$\tilde{{\cal{G}}}_{i}=U(1)$, since $\tilde{n}_{Vi} \leq 1$. The second equation,
however, has two solutions
\brr \tilde{n}_{Hi}=1, \qquad \tilde{n}_{Vi}=0, \qquad n_{Hi}=1
\label{eq:64}\err
and
\brr \tilde{n}_{Hi}=0, \qquad \tilde{n}_{Vi}=1, \qquad n_{Hi}=2 \,.
\label{eq:65}\err
These are illustrated in Figure 3 (a) and (b) respectively.
\begin{figure}
\begin{center}
\includegraphics[width=4in,angle=0]{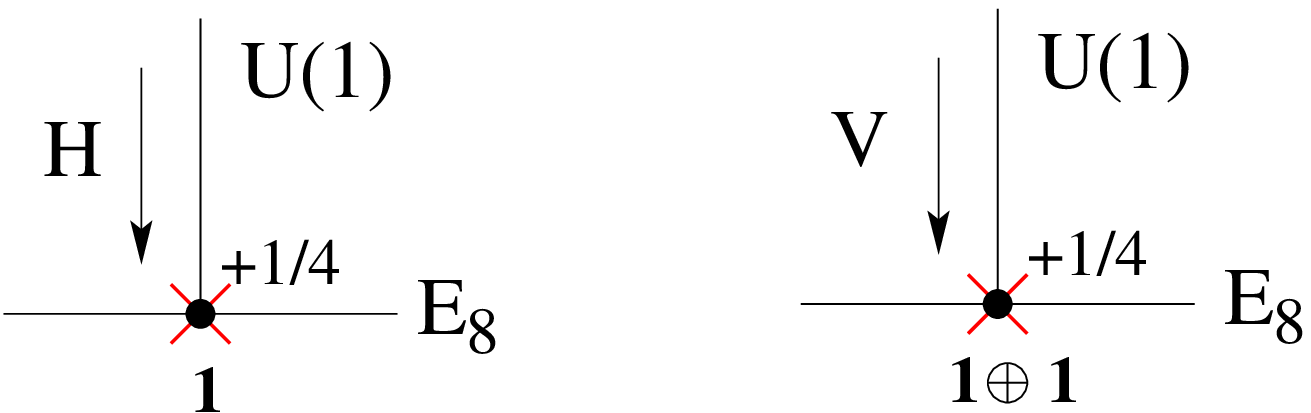}\\[.2in]
\parbox{6in}{Figures 3a and 3b: The two solutions with unbroken ten-dimensional $E_8$
and $n_T=1$ in which the seven-dimensional gauge group is $U(1)$.
These represent the cases where a fivebrane has wrapped the vertices
depicted in Figure 2. In each case, we have $\eta=1$, $\rho=1$ and
$g=+1/4$. The six-dimensional tensor multiplet is indicated by the
\includegraphics[width=.2in,angle=0]{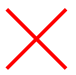} on the vertices.  }
\end{center}
\end{figure}

In either case, the anomaly (\ref{eq:53}) factors into an exact square given
by
\brr -\frac{3}{(2\pi)^{3}4!\,16}\,\bpl
     {\rm tr}\,R^{2}
     -\frac{1}{15}\,{\rm Tr}\,F_{i}^{2}\,\bpr^2 \,.
\label{eq:66}\err
The anomaly can now be cancelled by a Green-Schwarz mechanism on the
six-plane. First, one alters the Bianchi identity for the anti-self-dual tensor
in the twisted sector tensor multiplet
from $dH_{Ti}=0$, where $H_{Ti}$ is the tensor field strength three-form, to
\brr dH_{Ti}=\frac{1}{16\pi^{2}}
    (\,{\rm tr}\,R^{2}-\frac{1}{15}{\rm Tr}\,F_{i}^{2}\,) \,.
\label{eq:67}\err
Second, additional Chern-Simons terms are added to the action, one for each
six-plane. The required terms are
\brr S=\cdots-\frac{1}{64\pi}
    \sum_{i=1}^{32}\int \delta^{(5)}_{M_{i}^{6}}
    \wedge B_{Ti} \wedge (\,{\rm tr}\,R^{2}-\frac{1}{15}{\rm Tr}\,F_{i}^{2}\,) \,,
\label{eq:68}\err
where $B_{Ti}$ is the anti-self-dual tensor two-form on the $i$-th six-plane.
Using Bianchi identity (\ref{eq:67}), one can compute the variation of each
such term under Lorentz and gauge transformations. The result is a classical
anomaly that descends from an eight-form that {\it exactly cancels}
expression (\ref{eq:66}). The theory is now anomaly free.

Thus, we have demonstrated, within the context of an explicit orbifold
fixed plane intersection where the $\beta$ ${\bf Z \rm}_{2}$ projection to the
six-plane leaves $E_{8}$ unbroken, that {\it all local anomalies
can be cancelled}.
However, this cancellation requires that the intersecting
{\it seven-plane support a
twisted sector consisting of a $U(1)$ $N=1$ vector supermultiplet and an
associated Chern-Simons term}. This term is of the form (\ref{eq:49}) with
$\eta=1$ and $\rho=0$. The fact that $\rho=0$ in this context
follows directly from the
property that $E_{8}$ is unbroken by the $\beta$ projection.

We close this section by briefly presenting another possible orbifold fixed
plane intersection where local anomaly freedom {\it requires} that the
parameter $\rho$ in (\ref{eq:49}) be non-vanishing.
For this to be the case, we must allow
$E_{8}$ to be broken to a subgroup by the $\beta$ ${\bf Z \rm}_{2}$
projection to the six-plane. In this example, we take
\brr E_{8} \longrightarrow E_{7} \times SU(2) \,.
\label{eq:69}\err
This does not effect the $I_{8}(SG)_{i}$ contribution to the anomaly
eight-form given in (\ref{eq:20}), but does alter the untwisted gauge anomaly
from (\ref{eq:23}) to $I_{8}(E_{7} \times SU(2))_{i}$. The analysis of the
twisted sector spectrum on each six-plane remains identical to that discussed
above. As in the previous example, we can show that there can be no
twisted sector vector multiplets on any six-plane.  That is, the field
strengths ${\cal{F}}_{i}$ do not appear in the anomaly eight-form. Again, we
will assume that $G_{7i}=\tilde{{\cal{G}}}_{i}$ and limit our discussion to
the simplest case which, in this example, is
\brr \tilde{{\cal{G}}}_{i}=SU(2) \,.
\label{eq:70}\err
This seven-plane gauge group must be identified with the $SU(2)$ factor group
in $E_{7} \times SU(2)$.
\begin{figure}
\begin{center}
\includegraphics[width=3in,angle=0]{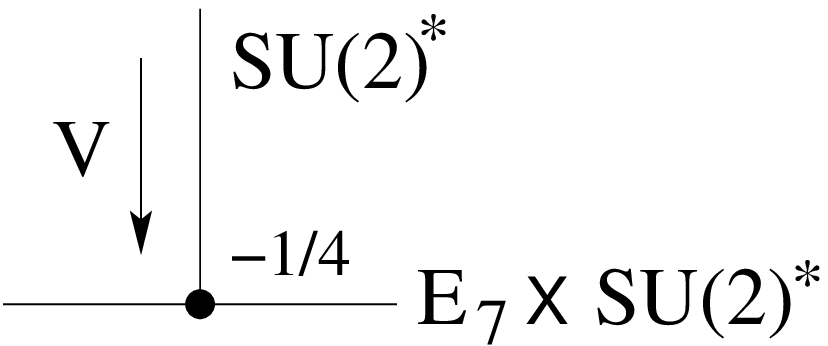}\\[.2in]
\parbox{6in}{Figure 4:  The solution with $n_T=0$ where
the ten-dimensional $E_8$ group is broken to $E_7\times SU(2)$.
This case requires a seven-dimensional gauge group $SU(2)$, which is
identified with the $SU(2)$ factor on the ten-plane.  This solution has
$\eta=\rho=1$ and $g=-1/4$.}
\end{center}
\end{figure}
Analysis of the cancellation of all anomalies at this fixed plane intersection
yields the following results. First of all, for there to be any solution, one
must take the $\rho$ parameter in (\ref{eq:50}) to be
\brr \rho=1 \,.
\label{eq:71}\err
For the case where
\brr n_{Ti}=0
\label{eq:72}\err
the complete anomaly will vanish identically if and only if
\brr g_{i}=-1/4, \qquad \eta=1
\label{eq:73}\err
and
\brr \tilde{n}_{Hi}=3, \qquad \tilde{n}_{Vi}=0, \qquad n_{Hi}=0
\label{eq:74}\err
where the $\tilde{n}_{Hi}$ hypermultiplets transform in the ${\bf (1,3) \rm}$
representation of $E_{7} \times SU(2)$.
This solution is illustrated in Figure 4.

For the case where
\brr n_{Ti}=1
\label{eq:75}\err
the anomaly eight-form will factorize into a complete square, and, hence, be
cancelled by a Green-Schwarz mechanism on the six-plane, if and only if
\brr g_{i}=3/4, \qquad \eta=1
\label{eq:76}\err
and
\brr \tilde{n}_{Hi}=3, \qquad \tilde{n}_{Vi}=0, \qquad n_{Hi}=1
\label{eq:77}\err
where the $\tilde{n}_{Hi}$ and $n_{Hi}$ hypermultiplets transform as ${\bf
(1,3) \rm}$ and ${\bf (1,1) \rm}$ respectively under $E_{7} \times SU(2)$.
This solution is illustrated in Figure 5.
\begin{figure}
\begin{center}
\includegraphics[width=3in,angle=0]{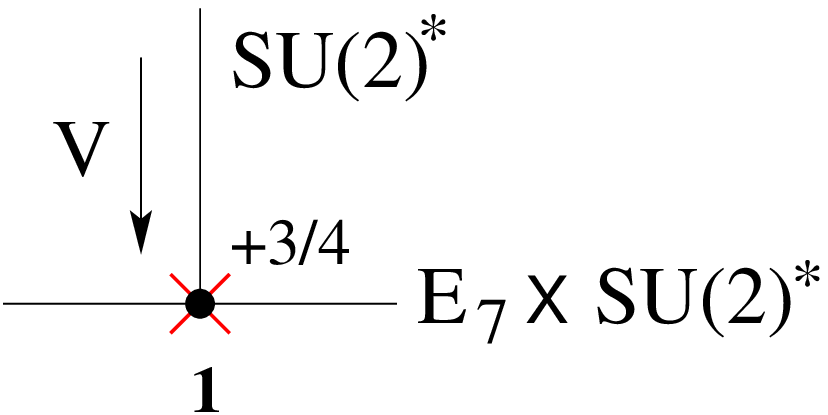}\\[.2in]
\parbox{6in}{Figure 5:
The solution with $n_T=1$ where
the ten-dimensional $E_8$ group is broken to $E_7\times SU(2)$.
This represents the case where a fivebrane has wrapped the vertex depicted
in Figure 4.  The solution in Figure 5 has
$\eta=\rho=1$ and $g=-1/4$.The six-dimensional tensor multiplet is indicated by the
\includegraphics[width=.2in,angle=0]{X.eps} on the vertex.  }
\end{center}
\end{figure}

Therefore, within the context of a second orbifold fixed plane intersection
where the $\beta$ ${\bf Z \rm}_{2} $ projection breaks $E_{8} \longrightarrow
E_{7} \times SU(2)$, {\it all local anomalies can be cancelled}. This
cancellation requires that the intersecting {\it seven-plane support a twisted
sector consisting of an $SU(2)$ adjoint representation of $N=1$ vector
supermultiplets}. In addition, {\it the seven-plane must support an
associated Chern-Simons term} of the form (\ref{eq:49}) with
$\eta=\rho=1$. \\[.2in]
{\bf Acknowledgements:}\\[.1in]
BAO would like to thank Jens Erler for intersting conversations.
Burt Ovrut is supported in part by the DOE under contract No.
DE-ACO2-76-ER-03071.


\begin{thebibliography}{99}

\bibitem{hw1}
P.Ho{\v r}ava and E.Witten,
{\it Heterotic and Type I String Dynamics from Eleven Dimensions},
Nucl.Phys. {\bf B}475 (1996) 94-114,
hep-th/9510209.
%
\bibitem{hw2}
P.Ho{\v r}ava and E.Witten,
{\it Eleven-Dimensional Supergravity on a Manifold with Boundary},
Nucl.Phys. {\bf B}460 (1996) 506-524,
hep-th/9603142.
%
\bibitem{losw1} A. Lukas, B.~A. Ovrut, K.~S. Stelle and D. Waldram,
    {\em The Universe as a Domain Wall},
    Phys.Rev. D59 (1999) 086001; {\em Heterotic M-theory in Five Dimensions},
    Nucl.Phys. B552 (1999) 246-290; A. Lukas, B.~A. Ovrut and D. Waldram,
    {\em Non-Standard Embedding and Five-Branes in Heterotic M-Theory},
    Phys.Rev. D59 (1999) 106005; R. Donagi, A. Lukas, B.~A. Ovrut and
    D. Waldram, {\em Non-Perturbative Vacua and Particle Physics in M-Theory},
    {\em JHEP} 9905 (1999) 018; {\em Holomorphic Vector Bundles and
    Non-Perturbative Vacua in M-Theory}, {\em JHEP} 9906 (1999) 034;
    A. Lukas, B. A. Ovrut and  D. Waldram, {\em
    Five--Branes and Supersymmetry Breaking in M--Theory},
    {\em JHEP} 9904 (1999) 009;
    R. Donagi, B. A. Ovrut and  D. Waldram, {\em
    Moduli Spaces of Fivebranes on Elliptic Calabi-Yau Threefolds},
    {\em JHEP} 9911 (1999) 030;
    R. Donagi, B. A. Ovrut, T. Pantev and  D. Waldram, {\em
    Standard Models from Heterotic M-theory},
    hep-th/9912208;
    R.Donagi, B.Ovrut, T.Pantev and D.Waldram,
    {\em Standard Model Bundles on Non-Simply Connected Calbi-Yau Threefolds},
    hep-th/0008008; R.Donagi, B.Ovrut, T.Pantev and D.Waldram,
    {\em Standard Model Bundles},
    math.AG/0008010; R.Donagi, B.Ovrut and D.Waldram,
    {\em Spectral Involutions on Rational Elliptic Surfaces},
    math.AG/0008011.
%
\bibitem{dasmuk}
K.Dasgupta and S.Mukhi,
{\it Orbifolds of {\it M}-theory},
Nucl.Phys. {\bf B}465 (1996) 399-412,
hep-th/9512196.
%
\bibitem{wittens5}
E.Witten,
{\it Five-branes and {\it M}-Theory on an Orbifold},
Nucl.Phys. {\bf B}463 (1996) 383-397,
hep-th/9512219.
%
\bibitem{mlo}
M.Faux, D.L{\"u}st and B.A.Ovrut,
{\it  Intersecting Orbifold Planes and Local Anomaly Cancellation
in {\it M}-theory},
Nucl.Phys. {\bf B}554 (1999) 437-483,
hep-th/9903028
%
\bibitem{phase}
M.Faux, D.L{\"u}st and B.A.Ovrut,
{\it Local Anomaly Cancellation, {\it M}-theory Orbifolds
and Phase-transitions},
hep-th/0005251
%
\bibitem{new}
M.Faux, D.L{\"u}st and B.A.Ovrut,
{\it An $M$-theory Perspective on Heterotic $K3$ orbifold Compactifications},
hep-th/0010087
%
\bibitem{bds}
Adel Bilal, Jean-Pierre Derendinger, Roger Sauser,
{\it  M-Theory on $S^1/Z_2$ : New Facts from a Careful Analysis},
hep-th/9912150.
%
\bibitem{ksty}
V. Kaplunovsky, J. Sonnenschein, S. Theisen, S. Yankielowicz,
{\it On the Duality between Perturbative Heterotic Orbifolds
and M-Theory on $T^4/Z_N$},
hep-th/9912144.
%
\bibitem{opp} B. A. Ovrut, T. Pantev and  J. Park,
{\it Small Instanton Transitions in Heterotic M-Theory},
hep-th/0001133.


\end{thebibliography}
\end{document}